# VOID SIZES AND CMBR FLUCTUATIONS

T. Piran
*Racah Institute for Physics, The Hebrew University, Jerusalem, Israel.*

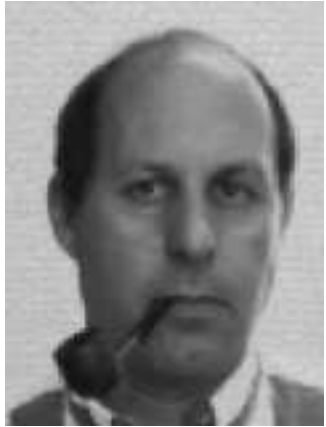

**Abstract**

Following Blumenthal *et al.* [1] and Piran *et al.* [2] I discuss a simple model for the gravitational formation of voids. This model enables us to estimate the primordial perturbation required on the scale of voids. A comparison of this condition with the constraints on the spectrum of fluctuations derived from CMBR observations reveals that gravitational growth of large voids require more power on the scale of $\approx$ 1500km/s than is generally accepted in CDM models. For a generic power law primordial spectrum $P(k) \propto k^n$ gravitational growth of voids requires $n \gtrsim 1.25$, which is marginally compatible with the observed CMBR fluctuations. If future measurements will show that the amplitude of fluctuations required to produce the voids is too large they will imply that the voids did not grow gravitationally and that galaxies do not trace matter on very large scale.

## 1  INTRODUCTION

Recent observations of the galaxy distribution strongly suggest that galaxies tend to lie on wall-like features which bound large empty regions, forming a closely packed network of voids [3, 4, 5, 6, 7]. This property seems to be generic since it is present in all the slices in the northern and southern hemisphere that have been completed by the ongoing CfA2 and SSRS2 surveys [6, 8]. Further evidence for this picture has been provided by the red-shift maps of the deep survey being conducted by [9]. Combined these surveys are qualitatively suggestive of a void-filled Universe. Gross estimates of the typical size and under-density suggest that the voids have a characteristic scale ranging from 2500 to 5000 km/s in diameter with a typical under-density estimated at about 20% of the mean (see [11], for the Boötes void). Currently there is no clear statistical measure for the properties of voids. Fig. 1 depicts a preliminary distribution of voids in the SSRS2 survey obtained using a new void search algorithm by El-Ad *et al.* [10]. The figure supports the qualitative picture described above.

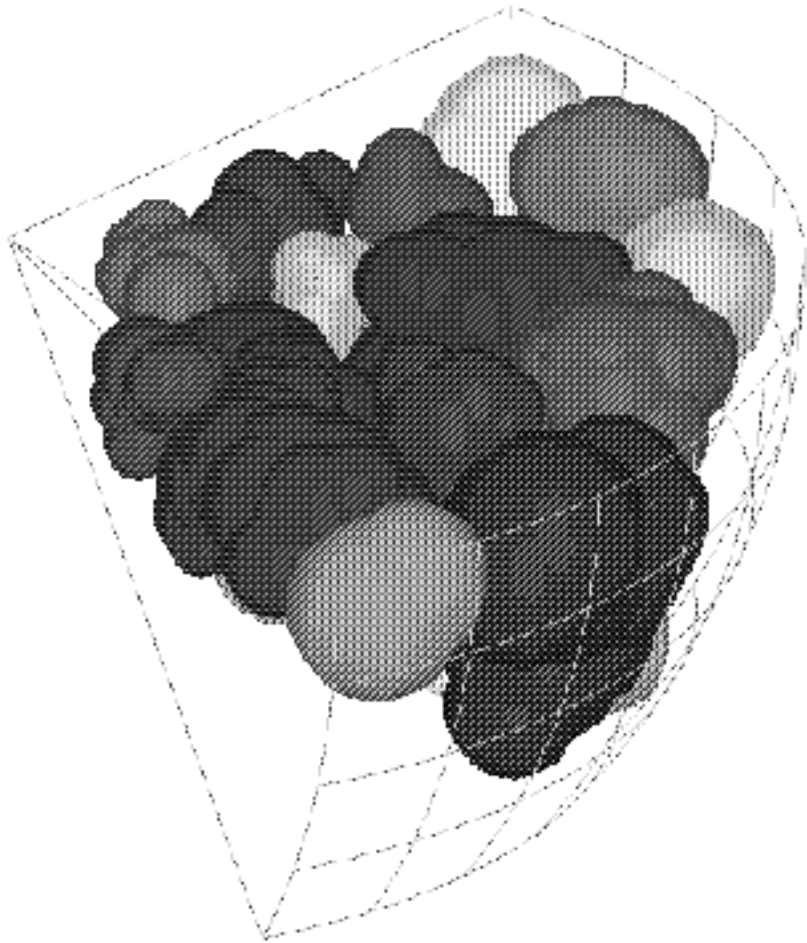

*Fig. 1: Voids in the SSRS2 survey [8]. The survey is in the south galactic hemisphere and its boundaries are: $20.9h < RA < 4.66 hours$ and declination $-40° < \delta < -2.5°$. The figure depicts a slice of constant declination in the range $-12.5° < \delta < -7°$ from [10]*

Following Blumenthal *et al.* [1] Dubinski *et al.* [12] and Piran *et al.* [2] I describe here a simple model for the gravitational formation of voids. According to this model voids grow gravitationally from negative density fluctuations which are now at the stage of shell crossing. The condition of shell crossing constraint the primordial fluctuation spectrum on the scale of voids. This criterion enables us to incorporate the existence of voids in the quest for the primordial spectrum of fluctuations. Alternatively, if one can show that this constraints cannot be satisfied by the primordial fluctuation spectrum we will have to conclude that light does not trace matter on the scale of voids.

## 2 GRAVITATIONAL GROWTH OF VOIDS

Consider, first, the growth of a single isolated spherical inverted top-hat void in an $\Omega = 1$ universe. We assume that at some initial time $t_i$ (e.g. at horizon crossing) there is an inverted top-hat density distribution; i.e., for $r \leq R_i$, $\rho = \rho_c(1 - \epsilon_i)$ where $\rho_c$ is the critical density and $\epsilon_i = 1 - \Omega_i$, and for $r > R_i$, $\rho = \rho_c$. Initially this perturbation grows linearly: the under-density increases with time while its size grows only with the expansion of the universe. Eventually the perturbation becomes non-linear and its size increases faster than the expansion of the universe. Finally at a time $t_{sc}$ given by $\epsilon_i(t_{sc}/t_i)^{2/3} \approx \delta_{crit} \approx 2.5$, (i.e. when $\epsilon_i(1 + z_i)/(1 + z_{sc}) \approx \delta_{crit}$) shell crossing begins (with the shell just inside the boundary being the first to cross). We define the instant of shell crossing as the moment at which the void forms. This definition might seem somewhat arbitrary. However, at $t_{sc}$ the co-moving radius $R_{sc}$ has grown and it is 1.7 times the initial co-moving radius $R_i$. The density in the void is then $1/1.7^3 \approx 0.2$ of the average density, which is the frequently quoted value for the galaxy under-density in the observed voids. The requirement for shell crossing may perhaps be too extreme, although it naturally leads to under-density values which are roughly comparable to observations. Relaxing this condition and requiring the under-density to be 30% of the mean (which corresponds to an expansion over the co-moving scale by a factor of 1.45) implies that the amplitude of the initial perturbation could be decreased by a factor of 1.5.

Despite the simplicity of the adopted model, its main features are supported by N-body experiments of simple configurations of several interacting voids carried out by Dubinski *et al.* [12] . In these simulations it was found that the condition for formation of an isolated void holds approximately, even for highly non–spherical perturbations and when other negative or positive perturbations are present. These numerical simulations also reveal that voids at shell–crossing are the most prominent being delineated by high-density contrast walls. Only tenuous traces of smaller scale voids, past the shell-crossing phase, are seen producing a distribution that roughly resembles that actually observed for the galaxies.

Two other properties are critical to our model. First, the largest voids at any epoch are those reaching shell crossing at that epoch, since after shell-crossing voids grow relatively slowly. Second, if the fluctuations are Gaussian-distributed then when rms negative fluctuations on a given scale reach shell crossing the corresponding voids will also occupy a large fraction of the volume of the universe, resembling a void-filled universe [12]. We use this fact to identify $\epsilon$ with $\sigma$, the rms mass fluctuation on the scale of voids. It might be that in a more realistic model $1.5\sigma$ or $2\sigma$ fluctuations will be sufficient to produce a void-filled universe. This introduces an additional uncertainty in our model which is difficult to estimate.

The quantity $\epsilon_i(1 + z_i)/(1 + z)$ equals to the amplitude of the perturbation $\delta M/M_L$ had it continued to grow *linearly* until today. It can be expressed in terms of the linear perturbation spectrum $P_L(k)$ as:

$$\left\langle \left(\frac{\delta M}{M}\right)^2_L (R) \right\rangle = \frac{1}{2\pi^2} \int_0^\infty dk \ k^2 P_L(k) \ \mathcal{W}^2(kR) \quad . \tag{1}$$

$\mathcal{W}(kR)$ is a window function, for which we use the Fourier transform of a Gaussian which has been normalized to have the same spatial volume as a top hat of radius $R$:

$$\mathcal{W}(x) = \exp[-\frac{1}{2}x^2(2/9\pi)^{1/3}] \quad . \tag{2}$$

This choice of a window function assures the convergence of the integrals in Eq. 1. For a given linear power spectrum we can equate $\delta M/M$ to $\delta_{crit}$ . This provides an equation for $R$ the scale of voids today. Alternatively, if the scale of voids is known then then this equality becomes an integral equation for $P_L(k)$.

# 3 LIMITS ON THE POWER-SPECTRUM FROM CMBR DATA

The same mass fluctuations that determine the gravitational growth of voids are also responsible for the CMBR fluctuations measured today. For a given form of the power-spectrum we normalize the power-spectrum using the observed CMBR fluctuations. This will determine the co-moving scale that satisfies the condition $\delta M/M = \delta_{crit}$ and in turn will predict the size of voids today, consistent with the CMBR fluctuations. This prediction should be compared with the observed scale of the voids.

The fluctuations in the CMBR on scales larger than one degree, which are of interest here, are dominated by the Sachs-Wolfe effect [13]. In a flat ($\Omega = 1$) universe, $C(\theta)$ can be expressed in terms of the current *linear* power spectrum, $P_L(k)$, as:

$$\frac{C(\theta)}{T^2} = \frac{H_o{}^4}{8\pi^2 c^4} \int_0^\infty \frac{dk}{k^2} P_L(k) \frac{\sin k\zeta}{k\zeta} \quad , \tag{3}$$

where

$$\zeta \equiv 2R_h \sin(\theta/2) \quad , \tag{4}$$

and $R_h = 2c/H_o$ is the horizon size.

For a given $C(\theta)$, Eq. 3 is another integral equation on $P_L(k)$. For a given functional shape for $P_L(k)$, such as the case for a CDM model, Eq. 3 determines the amplitude of the linear power spectrum $P_L(k)$.

# 4 VOIDS AND CMBR ANISOTROPY

Using the results of the preceding sections we can now examine the size of typical voids today which are consistent with the constrains on the power-spectrum imposed by the CMBR experiments. Following Blumenthal *et al.* [1] and Piran *et al.* [2] we choose a power spectrum, $P_L(k)$, of the form $P_L(k) = A_n k^n$. This might be justified since we are interested in extrapolation from the scale of COBE measurements to the scale of the voids. We discuss, later, qualitatively the implication of replacing this form by the linear CDM power spectrum. For each $n$ we integrate Eq. 3 and define, $F_n(\theta)$ such that :

$$\frac{C(\theta)}{T^2} \equiv \frac{2}{\pi^2} \left(\frac{H_o}{2c}\right)^{n+3} A_n F(\theta) \quad . \tag{5}$$

Using the COBE data [14] we then determine the appropriate normalization constant, $A_n$ and integrate Eq. 1 to obtain an estimate of $\delta M/M_L$ as a function of the power index $n$:

$$\frac{\delta M}{M}_L(v) = \sqrt{2^n \left(\frac{9\pi}{2}\right)^{(n+3)/6} \Gamma\left(\frac{n+3}{2}\right)} \sqrt{\left\langle\frac{C_{obs}}{T^2 F_n}\right\rangle} \left(\frac{c}{v}\right)^{(n+3)/2} \quad . \tag{6}$$

We have used the velocity, $v = H_0 R$, instead of the radius, $R$, to obtain a formula that is independent of the Hubble constant. We also remind the reader that since at shell crossing $R_{sc} \approx 1.7 R_i$, a typical void with a diameter of $2R_{sc}$ today corresponds to an initial perturbation with a co-moving radius of $R_i$.

The condition imposed by the existence of voids that evolve gravitationally from rms fluctuations can be expressed by equating Eq. 6 to the value of $\delta_{crit} = 2.7$ discussed in section II ($\delta_{crit}$ could be as low as 1.8 (if we consider under-density of 30% in the voids or if $1.5\sigma$ rather than $1\sigma$ fluctuations produce the typical voids). Piran *et al.* [2] calculated $\delta M/M_L$ as a function of

$v_f = 3.4 \times v$, where $v$ corresponds to the co-moving size in km/s of the unevolved primordial perturbation for different values of $n$. These results are summarized in Fig. 2 and in Table 1. They find that in order to form typical voids of 5000 km/s in diameter we must consider a power-spectrum with $n \approx 1.25$. This could be the case if the primordial spectrum is steeper than expected from inflation.

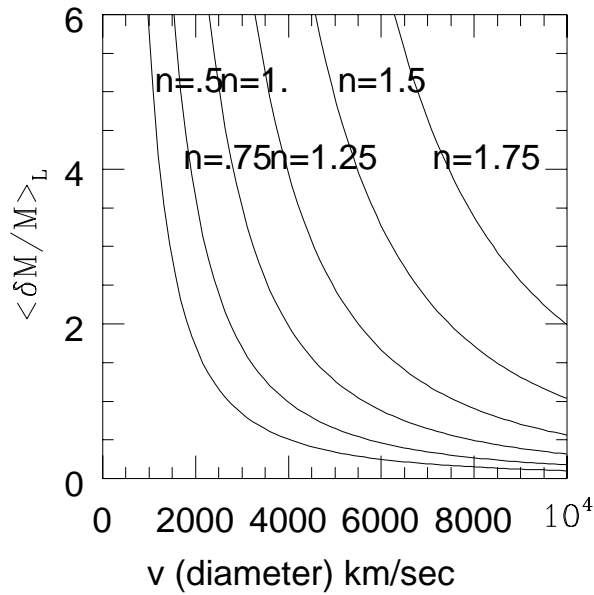

Fig. 2: Predicted $\delta M/M_L$ vs. the void diameter (measured in km/s ) for various power laws spectrum, whose amplitude fits COBE, from [2].

**Table 1.** I Comparison of CMBR and voids for different spectral indices [2].

| $n$ | $\sqrt{\left\langle \frac{\tilde{C}_{obs}}{T^2 F} \right\rangle}$ | $\frac{\delta M}{M_L}$ [1] | $v$ [2] | $\sqrt{C_{1.2}(0)}$ [3] |
|---|---|---|---|---|
| .50 | $1.3 \cdot 10^{-5}$ | 0.34 | 1500 | $1.3 \cdot 10^{-5}$ |
| .75 | $.1 \cdot 10^{-5}$ | 0.63 | 2400 | $1.4 \cdot 10^{-5}$ |
| 1.00 | $.89 \cdot 10^{-5}$ | 1.2 | 3400 | $1.6 \cdot 10^{-5}$ |
| 1.25 | $.74 \cdot 10^{-5}$ | 2.4 | 4800 | $1.9 \cdot 10^{-5}$ |
| 1.50 | $.65 \cdot 10^{-5}$ | 4.7 | 6500 | $2.4 \cdot 10^{-5}$ |

1) $\delta M/M_L$ on a scale that corresponds to voids with a diameter of $v = 5000 km/sec$ today.
2) Diameter of the voids (using $\delta M/M_L = 2.7$) in $km/sec$.
3) $\sqrt{C(0)}$ convolved with a FWHM beam of $1.2^0$. This should be compared with the $2\sigma$ upper limit of Gaier et al. [15] of $1.4 \cdot 10^{-5}$.

For a Harrison-Zel'dovich spectrum ($n = 1$), consistent with the COBE observations and the South Pole anisotropy upper-limit, $\delta M/M_L$ is $\approx 1.2$ at the scale of 5000 km/s , and thus such voids are not expected to be typical. The typical diameter of volume-filling voids should be in this case $\approx 3500$ km/s . Voids as large as 5000 km/s could still be formed from $2\sigma$ fluctuations and they would represent the high-end tail of the distribution of void sizes.

Of course, some of the difficulties of forming large voids could be alleviated if we consider the possibility that voids are essentially formed when the under-density is of the order of 30%. In this case the required amplitudes are decreased by a factor of 1.5 and voids with diameters of

4200 km/s can form from 1$\sigma$ fluctuations for $\delta_{crit}$ =2.7, (3500 km/s for $\delta_{crit}$ =4.5) for a Harrison-Zel'dovich spectrum. We note, however, that in this case we would not expect them to fill in the volume. We also expect the voids to have shallower profiles and less sharp boundaries. Unfortunately, the available data is far from adequate to address these details.

Alternatively, if 1.5$\sigma$ fluctuations are sufficient to produce a volume filled Universe, (or if the voids are not volume filling) this will also reduce by a similar factor the required amplitude for formation of voids and 4200 km/s voids could form with a $n = 1$ spectrum. However, one has to relax both assumptions – i.e. to require that the observed voids are shallower and that they are produced by 1.5$\sigma$ events – to explain the gravitational formation of 5000 km/s voids with such a spectrum.

Piran et al. [2] have considered only generic power-law spectra, but this approach can also be used to predict the size of typical voids for any specific model of primordial fluctuations. If we consider CDM models we should replace the power-law spectrum by another function, such as the one proposed by Efstathiou, et al. [17]:

$$P(k) = \frac{Ak^n}{1 + 1.7k/(\Omega h) + 9[k/(\Omega h)]^{3/2} + [k/(\Omega h)]^2} \qquad (7)$$

where $\Omega$ is the closure parameter and $h$ the Hubble constant measured in $100 km/sec/Mpc$. This function has the same power on large scale as a pure power law but less power on smaller scales. Since COBE measurements are on large scale, substitution of Eq. 7 to Eq. 3 will give effectively the same normalization. However, its substitution to Eq. 1, will give lower $\delta M/M_L$ on a given scale $R$. Hence the overall effect of replacing the generic power law spectrum by a more realistic CDM model is to increase the required power law (which appears also in the CDM perturbation spectrum. Recent calculations along this line by Lecar and Gregory [16] show that the required index increases to $n = 1.5$ (with the standard assumptions described above). This result is mildly in conflict with smaller scale ($\approx 1^o$) CMBR measurements, whose comparison with the COBE data suggest that $1 < n < 1.25$ [18]. With the standard parameters of CDM: a Harrison-Zel'dovich spectrum $n = 1$, $h = 0.65$, and the CDM transfer function (Eq. 7), normalized by the two-year COBE results, they shows that the amplitude required to form 5000km/s voids gravitationally is achieved only for 3-sigma events.

Lecar & Gregory [16] have also examined an open Universe with cosmological constant such that $\Omega = 1$ but $\Omega_{matter} < 1$. In this universe $R_h$, (which is proportional to $\Omega^{-1}$) increases and this results in a much larger separation between the scale of COBE ($\geq 10^o$) and the scale of voids ($5000km/sec$). At the same time the perturbations gradually stop growing (after $1 + z \approx \Omega_{matter}^{-1}$). The combined result of both effects is that the required power law index, $n$, remains practically unchanged. The examination of a $\Omega < 1$ universe is much more complicated (see for discussion of CMBR fluctuations [19]) and it is under consideration now.

## 5 DISCUSSION

I have presented here a quantitative model for the incorporation of the information on the existence of voids to the analysis of the primordial perturbation spectrum. Our basic result is that the "common" existence of voids on scales ranging from 2500 to 5000 km/s requires, if they formed gravitationally, more power in the range 800 to 1500 km/s than predicted from the standard unbiased CDM model. This implies that non-linear effects should be important in the evolution of perturbations on these scales. This contrasts with the conventional idea that scales beyond about 8 $h^{-1}$ Mpc are still in the linear regime. With our standard assumptions: $\delta_{crit}$ =2.7, 1$\sigma$ fluctuations produce the observed voids ( with an under-density of 20%) and

$P(k) \propto k^n$, a power law index of $n = 1.25$ is required to produced a 5000 km/s void filled Universe and that a Harrison-Zel'dovich, ($n = 1$), spectrum can produce gravitationally, only 3500 km/s voids. Similar considerations constrain the largest possible void to about 6000 km/s in diameter. Voids as large as those suggested by Broadhurst *et al.* [20] cannot form gravitationally and they cannot correspond to voids in the matter distribution.

We expect that in our model there will be a relationship between the abundance of voids, whose origin is negative density fluctuations and rich clusters that originate from positive ones. For a void-filled universe with a scale of 5000 km/s we expect (assuming that rich cluster form from positive density fluctuations with the same initial amplitude as the negative perturbations that form voids) the abundance of dark matter clusters with masses of $5 \times 10^{15} M_\odot$ to be of the order of $8 \times 10^{-6} H^{-3}/\mathrm{Mpc}^3$, corresponding roughly to galaxy clusters of $5 \times 10^{14} M_\odot$. This should be compared with an Abell richness 1 cluster density of one per $(55 h^{-1} Mpc)^3$ and with the density of Abell clusters of richness 2, one per $(95 h^{-1} Mpc)^3$ ([21]). It seems that our model predicts more rich clusters than observed. However, given all the uncertainties involved in the model and the data available for clusters we cannot discard the possibility of non-linearity on the basis of the existing data on cluster abundance. Specifically, if voids form from $1.5\sigma$ fluctuations the symmetry between positive and negative fluctuations is broken, reducing the predicted cluster density. Alternatively, it is possible that the primordial perturbation spectrum is not Gaussian.

The basic assumption of our model is that the "voidy" nature of the galaxy distribution is a natural consequence of negative rms amplitude fluctuations which grow in size during the evolution of the universe producing an essentially void-filled distribution. Although the details of our model are uncertain and the information on voids is still sketchy, our formalism can be used to quantitatively assess the likelihood of a given model of structure formation to generate voids. The underlying assumption of our approach is that voids grow gravitationally, that light traces matter at large scales and therefore the large voids in the galaxy distribution correspond to real voids in the matter distribution. If further evidence will suggest that there is indeed contradiction between the power required for gravitational formation of voids and other observations this will imply that *the voids in the galaxy distribution do not trace the matter distribution.* This will further imply that the observed power spectrum measured from large scale galaxy surveys does not reflect the underlying matter power spectrum on the scale of the voids.

It is important to stress that the simple model in which galaxies do not form in under-dense regions, and consequently, the under-density in the the dark matter is greatly enhanced in the galaxy distribution is not sufficient to resolve the large voids problem. In this model we would expect that within the observed voids there will appear marked "walls" of galaxies, which reflect the over-density on smaller scales that exists in the dark matter. Put differently, in order for a biasing scheme of this kind to create a large void which does not follow the dark matter distribution, it will also have to erase structures corresponding to large amplitude dark matter perturbations within the void. A simple interpretation of this result is that the voids do not result from a gravitational phenomenon.

In summary there are two potential implications to our analysis. If this simple model is valid then either the power spectrum of primordial perturbation has a slope $n \gtrsim 1.25$ or the voids in the galaxy distribution do not result from a gravitational phenomenon. The former possibility is incompatible with simple inflationary models (but it is possible with more complicated ones that allow for two scalar fields). The second possibility means that galaxy distribution on scales of $\gtrsim 3000$km/s does not follow the dark matter distribution.

**Acknowledgements.** I would like to thanks G. R. Blumenthal, L. da Costa, H. El -Ad, S. Eyal, D. Goldwirth, D. Langlois and M. Lecar for many helpful discussions.

# References

[1] Blumenthal, G. R., da Costa, L., Goldwirth, D. S., Lecar, M. & Piran, T., (1992), Ap. J.. **388**, 234.

[2] Piran, T., Lecar, M., Goldwirth, D. S., da Costa, L. N., & Blumenthal, G. R., (1993), MNRAS, **265**, 681.

[3] Kirshner, R. P., Oemler, A. Jr., Schechter, P. L., & Shectman, S. A., (1981) Ap. J. Lett. **248**, L57.

[4] Kirshner, R. P., Oemler, A. Jr., Schechter, P. L., & Shectman, S. A., (1983) AJ, 88, 1285.

[5] De Lapparent, V., Geller, M. J., & Huchra, J. P., (1986) Ap. J. Lett. **302**, L1.

[6] Geller, M. J., & Huchra, J. P., (1989) Science **246**, 897

[7] da Costa, L. N., Pellegrini, P. S., Sargent, W. L. W., Tonry, J., Davis, M., Meiksin, A., Latham, D. W., Menzies, J. W., & Coulson, I. A., (1988) Ap. J. **327**, 544.

[8] da Costa, L. N., Geller, M. J., Pellegrini, P. S., Latham, D. W., Fairall, A. P., Marzke, R. O., Willmer, C. N. A., Huchra, J. P., Calderon, J. H., Ramella, M. & Kurtz, M. J., (1994) Ap. J. **424** L1

[9] Schectman, S. A., Schechter, P. L., Oemler, A. A., Tucker, D., Kirshner, R. P. & Lin, H., (1992) to appear in Clusters and Superclusters of Galaxies, ed A. C. Fabian.

[10] El-Ad, H., Piran, T. & da Costa, L. N., (1995) in preparation.

[11] Dey, A., Strauss, M. A., & Huchra, J., (1990) AP, **99**, 463.

[12] Dubinski, J., da Costa, L. N., Goldwirth, D. S., Lecar, M. & Piran, T., (1993) Ap. J. **410**, 458.

[13] Sachs, R. K. & Wolfe, A. M., (1967), Ap. J. **147**, 73.

[14] Smoot, G. F. *et al.,* (1992) Ap. J. Lett. **396**, L1.

[15] Gaier, T. *et al.,* (1992) Ap. J. Lett. **398**, L1.

[16] Lecar, M., & Gregory, B., (1995) submitted to AP. J.

[17] Efstathiou, G., Bond, J. R., & White, D. M., (1992) MNRAS, **258**, 1p.

[18] Dodelson, S. & Kosowsky, A., (1994) preprint FNAL-pub94/357-A, Astro-Ph-9410081.

[19] Kamionkowski, M. & Spergel, D. N., (1994) Ap. J., **432**, 7.

[20] Broadhurst, T. J., Ellis, R. S., Koo, D. C., & Szalay, A. S., (1990) Nature, **343**, 726.

[21] Bahcall, N. & Cen, R., (1992) Ap. J. Lett., **398**, L81.